\documentclass[pre, twocolumn,showpacs]{revtex4}
\usepackage{amsmath}
\usepackage{amssymb}
\usepackage{graphicx}
\usepackage{algpseudocode, algorithm}
\usepackage{color}

\begin{document}


\title{Decentralized Routing on Spatial Networks with Stochastic Edge Weights}



\author{Till Hoffmann}
\affiliation{Astrophysics Group, Imperial College London, London, United Kingdom}

\author{Renaud Lambiotte}
\affiliation{Naxys, University of Namur, Namur, Belgium and Department of Mathematics, University of Namur, Namur, Belgium}

\author{Mason A. Porter}
\affiliation{Oxford Centre for Industrial and Applied Mathematics, Mathematical Institute, University of Oxford, Oxford, United Kingdom and CABDyN Complexity Centre, University of Oxford, Oxford, United Kingdom}


\pacs{89.75.Hc, 89.40.-a, 84.40.Ua, 89.20.Hh}



\begin{abstract}

We investigate algorithms to find short paths in spatial networks with stochastic edge weights.  Our formulation of the problem of finding short paths differs from traditional formulations because we specifically do not make two of the usual simplifying assumptions: (1) we allow edge weights to be stochastic rather than deterministic; and (2) we do not assume that global knowledge of a network is available. We develop a decentralized routing algorithm that provides en route guidance for travelers on a spatial network with stochastic edge weights without the need to rely on global knowledge about the network.  To guide a traveler, our algorithm uses an estimation function that evaluates cumulative arrival probability distributions based on distances between pairs of nodes. The estimation function carries a notion of proximity between nodes and thereby enables routing without global knowledge.  In testing our decentralized algorithm, we define a criterion that makes it possible to discriminate among arrival probability distributions, and we test our algorithm and this criterion using both synthetic and real networks.

\end{abstract}


\maketitle


\section{Introduction} 

One of the most important aspects of many networks is their navigability \cite{Clauset2003How-Do-Networks,Boguna2009Navigability,Erola2011Structural}, and it is often important to find short paths between pairs of nodes in a network. For example, sending packages across the Internet, attempting to spread ideas through social networks, and transporting people or goods cheaply and quickly all require the ability to find paths with a small number of steps and/or a low cost \cite{Newman2010Networks}. 
Assuming that network topology and the cost of making a step is known, such paths can be found easily \cite{Dijkstra1959A-note}. Unfortunately, complete knowledge of network topology (and edge weights) is often unavailable or constitutes an insurmountable overhead \cite{Milgram1967The-small,Berman1998On-line,Peleg1989A-trade-off,Krioukov2007On-compact}. 

Despite the aforementioned difficulties, there is empirical evidence that some networks can be navigated by using only local information \footnote{Our initial motivation for studying our formulation of this problem arose when one of the authors accidentally left his umbrella at University of Bath and had to find a way to get it back to Oxford without further travel on his part.}. A well-known example is Milgram's small world experiments \cite{Milgram1967The-small}, which demonstrated that short paths between individuals in social networks exist and that individuals are able to navigate networks without global knowledge of network topology. This observation was put on solid theoretical ground more than 30 years later by Kleinberg \cite{Kleinberg2000Navigation}, who showed that one can find short paths between nodes via decentralized algorithms in certain types of spatially-embedded networks \cite{barth2011}.  This work has led to both theoretical and numerical studies of routing with limited information \cite{Newman2010Networks,Rosvall2005Navigating} as well as investigations of the importance of embedding a network in space when developing routing algorithms  \cite{Liben-Nowell2005Geographic, Aldous2010Connected,rio10,Lee2011Pathlength,Lee2012Geometric}. 
 
An important limitation of the above findings is their assumption that the cost of making a step is deterministic.  In many situations, it is much more appropriate to model 
the cost as a random variable. For instance, varying levels of traffic on networks \cite{Noland2002Travel,Fan2005Arriving,Nie2009Shortest} make it unsuitable to model such costs deterministically. The aim of the present article is to address this important limitation and to develop a decentralized algorithm for routing in networks with stochastic edge weights.

The remainder of this article is organized as follows.  We discuss the deterministic shortest path problem in Section \ref{dspp}, and we discuss a stochastic version of this problem in Section \ref{sspp}.  In Section \ref{crit}, we discuss criteria for measuring the quality of a path in a network.  In Section \ref{adapt}, we discuss an adaptive algorithm that will be helpful for trying to solve our stochastic shortest path problem.  We present the notion of an estimation function in Section \ref{sec:estimation}, and we discuss our new decentralised routing algorithm in Section \ref{sec:decentralized}.  We show the results of simulations on a synthetic network in Section \ref{sec:simulation} and on a real network in Section \ref{chicago}.  We discuss our results further in Section \ref{add} and conclude in Section \ref{conc}.  We present pseudocode for our decentralised algorithm and discuss additional technical details in appendices.


\section{Deterministic Shortest-Path Problem (DSPP)} \label{dspp}

A \emph{network} (or \emph{graph}) $G$ consists of a set $N$ of \emph{nodes} labeled by the indices $\left\{ i_{1},\ldots,i_{n}\right\}$ (and with cardinality $|N| = n$) and a set $E$ of \emph{edges} (with $|E| = m$) labeled by ordered pairs of indices $\left(i,j\right)$ that which indicate that there is a directed edge from node $j$ to node $i$. We associate a \emph{weight} $T_{ij}$, which represents a cost or travel time, with each edge $\left(i,j\right)$. A \emph{path} $\ell$ with $k$ steps is a sequence of $k+1$ nodes $\ell=\left\{ i_{1},\ldots,i_{k+1}\right\} $ that are connected to one another via edges. Note that we do not require a path to be ``simple," so a node can occur multiple times in a path.  See Ref.~\cite{Nie2006} for a discussion of cycles in paths that arise from adaptive routing.

The weight $T_\ell$ of a path $\ell$ is given by the sum of the weights of its constituent edges: 
\[
	T_{\ell}=\sum_{j=1}^{k}T_{i_{j}i_{j+1}}\,.
\]
The shortest-path problem (SPP) aims to determine the path of smallest total weight from an \emph{origin} node to a \emph{target} (or \emph{destination}) node.  In the DSPP, each edge weight $T_{ij}$ is deterministic, and a path with minimal total weight is called \emph{optimal}.


\section{Stochastic Shortest-Path Problem (SSSP)} \label{sspp}

Non-deterministic travel times are a typical feature of transportation networks \cite{Noland2002Travel}. Because this is our motivating example, we use the terms ``time" and ``weight" interchangeably.  

To define an SSSP, we let the weights $T_{ij}$ be real-valued random variables that are distributed according to a probability distribution function (PDF) with probabilities $p_{ij}$ \cite{Frank1969Shortest, Loui1983Optimal, Eiger1985Path, Fan2005Arriving, Nie2009Shortest}. In our SSSP formulation, we make three assumptions: (1) the random edge weights are independent of each other; (2) the PDFs do not change during the routing process; and (3) the weight incurred by traversing an edge becomes known upon completion of the step. (For example, the time taken to travel a road is known once the next junction is reached.)  With these assumptions, it follows that the PDF for the weight $T_{\ell}$ of a path $\ell$ to have a value $t$ is given by the convolution of the PDFs of the weights associated with the path \cite{Hoel1971Introduction}:
\begin{equation}
	p_{\ell}\left(t\right)=\left(\mathop{\ast}_{j=1}^{k}p_{i_{j}i_{j+1}}\right)\left(t\right)\,,\label{eq:path-convolution}
\end{equation}
where the right-hand side denotes $k$ consecutive convolutions. The probability to traverse the path $\ell$ and incur a weight $T_{\ell}\leq t$ is given by the cumulative distribution function (CDF)
\[
	U_{\ell}\left(t\right)=\int_{0}^{t}dt'\, p_{\ell}\left(t'\right)\,.
\]


\section{Criteria} \label{crit}

Because the edge weights are now random variables, we need to reconsider the concept of an optimal path. In particular, there is no longer a unique concept of optimality. For example, Frank \cite{Frank1969Shortest} defined a path to be optimal if its CDF surpasses a threshold $\theta$ within the shortest time, whereas Fan et al. \cite{Fan2005Arriving} suggested maximizing the CDF for a given time budget $\tau$. Each of these criteria has a regime in which it outperforms the other. If the budget $\tau$ available to a traveler is large, then there are many paths that result in almost certain arrival at the desired target; that is, $U_{\ell'}\left(\tau\right)\approx1$ for many paths $\ell'$. In this regime, the paths are virtually indistinguishable using Fan et al.'s criterion. However, Frank's criterion can easily identify the path that it deems to be optimal. In contrast, if the budget is small, then arrival at the target within the budget is unlikely; that is, $U_{\ell''}\left(\tau\right)\ll1$ for all paths $\ell''$. In this case, Frank's criterion is not helpful because there does not exist a path whose CDF surpasses the threshold. However, Fan et al.'s criterion can identify the path with the maximal CDF for the given budget. 

\begin{figure}
\begin{centering}
	\includegraphics{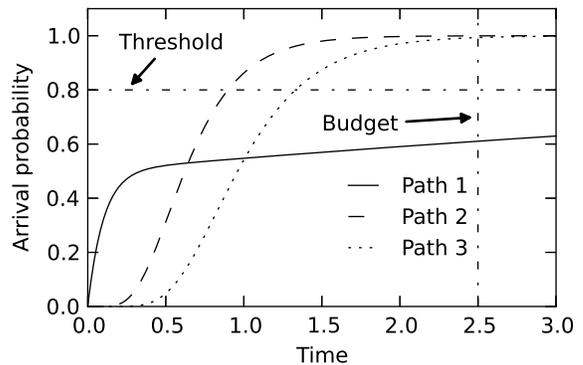}
\end{centering}
\caption{Comparison of path optimality criteria using CDFs of three paths. Fan et al.'s criterion prefers paths 2 and 3 to path 1 but cannot discriminate between the CDFs of paths 2 and 3. Frank's criterion prefers path 2 to path 3 but cannot be applied to path 1. The joint criterion is applicable to all CDFs and selects path 2 as the optimal one.}
\label{fig:criteria}
\end{figure}

We define a joint criterion that takes advantage of both aforementioned criteria. If there are paths whose CDFs surpass a threshold $\theta$ within the budget $\tau$, then we choose a path according to Frank's criterion. Otherwise, we choose a path according to Fan et al.'s criterion. In Fig.~\ref{fig:criteria}, we illustrate the differences between the criteria.

\section{An Adaptive Algorithm} \label{adapt}

Even finding approximate solutions to an SSPP is challenging. An interesting approach was proposed by Fan et al., who utilized an adaptive algorithm that evaluates the available information before each step and accounts for the consequences of previous decisions \cite{Fan2005Arriving}. They proposed building a routing table by considering the maximal probability to reach the target $r$ from all other nodes $i\in N\backslash\left\{r\right\} $. This amounts to solving the following set of coupled nonlinear integral equations:
\begin{align}
	u_{i}\left(t\right) &= \max_{j\in J_{i}}\left[\int_{0}^{t}p_{ij}\left(t'\right)u_{j}\left(t-t'\right)\, dt'\right]\,,\label{eq:fan-convolution}\\
	u_{r}\left(t\right) &= 1\,,
\end{align}
where $J_{i}$ is the set of neighbors of $i$ and $u_{i}\left(t\right)$ is the probability to arrive at node $r$ starting from node $i$ with a total travel time that is no longer than $t$. The node $q_{i}\left(t\right)$ that one should choose to attain the maximal arrival probability is 
\begin{equation}
	q_{i}\left(t\right)=\mathop{\arg\max}_{j\in J_{i}}\left[\int_{0}^{t}p_{ij}\left(t'\right)u_{j}\left(t-t'\right)\, dt'\right]\,.\label{eq:dominant-node}
\end{equation}

One cannot find analytical solutions to Eqs.~(\ref{eq:fan-convolution})\textendash{}(\ref{eq:dominant-node}) in general, but one can approximate the CDF $u_{i}\left(t\right)$ using the iterative sequence
\begin{align}
	v_{i}^{s+1}\left(t\right) & =\max_{j\in J_{i}}\left[\int_{0}^{t}p_{ij}\left(t'\right)v_{j}^{s}\left(t-t'\right)\, dt'\right]\,,\label{eq:fan-recursion}\\
	v_{r}^{s+1}\left(t\right) & =1\,, \nonumber 
\end{align}
with index $s$ and initial conditions
\[
	v_{i}^{0}\left(t\right)=0\,,\qquad\forall\, 
	i\in N\backslash\{r\}\,.
\]

The sequences $\{v_{i}^{s}\left(t\right)\}$ give lower bounds for the true CDFs. One obtains upper bounds by using the sequences $\{w_{i}^{s}\left(t\right)\}$ with the same recursion relation (\ref{eq:fan-recursion}) but with different initial conditions \cite{Fan2006Optimal}:
\[
	w_{i}^{0}\left(t\right)=1\,,\qquad \forall\, 
	i\in N\,.
\]
In our numerical implementation, we demand that the sequences converge to within a numerical tolerance $\epsilon$ for all $t$. That is, we require that 
\[
	w_{i}^{s}\left(t\right)-v_{i}^{s}\left(t\right)<\epsilon\,, \qquad \forall\, i\in N
\]
for sufficiently large $s$.


\section{Estimation Function}
\label{sec:estimation}

Fan et al.'s algorithm is centralized, because it requires knowledge of the entire network topology and all edge-weight distributions. To build a decentralized algorithm, we define an \emph{estimation function} $f\left(i,j;t\right)$, which gauges the arrival probability from node $j$ to node $i$ within time $t$. Such a function carries a notion of proximity between nodes, and we will use it to guide travelers on a network. We define an estimation function using the following four steps.

First, we embed the network under consideration in a metric space by defining a distance measure $d: N\times N\rightarrow \mathbb{R}$ for any pair of nodes. We use the shorthand notation $d_{ij}\equiv d(i,j)$ to denote the \emph{distance} between any pair of nodes $i$ and $j$ in the metric space. The \emph{length} of an edge is equal to the distance between a pair of nodes with a direct connection (i.e., an edge) between them. Note that the length of an edge (e.g., the length of a road) is different from its weight (e.g., the travel time).

Second, we define the \emph{network distance} \emph{$g_{ij}$} as the shortest distance between nodes if travelers are restricted to move along edges. Note that the network distance is distinct from the travel time. We assume that network distance between nodes $i$ and $j$ can be estimated from the (metric) distance between $i$ and $j$. That is, we assume that there exists a function $h$ such that $g_{ij}\approx h\left(d_{ij}\right)$. Such an assumption is implicit in all decentralized algorithms using a metric for guidance. 

Third, we note that the expected number of steps necessary to reach node $i$ from node $j$ is 
\begin{align*}
	\bar{k}_{ij} & =\left\lceil \frac{g_{ij}}{\lambda}\right\rceil \approx\left\lceil \frac{h\left(d_{ij}\right)}{\lambda}\right\rceil\,,
\end{align*}
where $\lambda$ denotes a characteristic edge length of a network and $\lceil x\rceil$ (which is called the ``ceiling" of $x$) denotes the smallest integer that is at least as large as $x$. 

Fourth, we assume that the weight $\bar{t}$ incurred by making a step towards the target is representative of the network and has a PDF of $\bar{p}(t)$. We estimate the weight of the unknown path $\bar{\ell}$ from $j$ to $i$ to be 
\[
	t_{\bar{\ell}}=\sum_{k=1}^{\bar{k}_{ij}}\bar{t}\,.
\]

Using Eq.~(\ref{eq:path-convolution}), we obtain an estimate
\[
	f\left(i,j;t\right)=
	\begin{cases}
	{\displaystyle \int_{0}^{t}dt'\,\left(\mathop{\ast}\limits _{k=1}^{\bar{k}_{ij}}\bar{p}\right)\left(t'\right)}\,, & \text{if }i\neq j,\\
	1\,, & \text{if }i=j\,.
	\end{cases}
\]
of the CDF. We thereby use physical distance to evaluate the number of steps between two nodes, and we assume that the random weight associated with each edge is uncorrelated with the length of the edge. In Section \ref{sec:simulation}, we discuss different choices for the characteristic edge length $\lambda$ and the characteristic distribution $\bar{p}(t)$. In Appendix \ref{app:conv}, we show that the order of carrying out mixtures and convolutions is irrelevant. 


\section{Decentralized Algorithm}
\label{sec:decentralized}

Our algorithm explores a network by using local information, and it chooses a locally optimal node according to one of the criteria discussed in Section \ref{crit}. The visited nodes $N_{\mathrm{V}}$ and the ``frontier nodes" $N_{\mathrm{F}}$ constitute a known subgraph $G_{\mathrm{K}}$ (i.e., the parts of the graph that the traveler has discovered). Frontier nodes are neighbors of visited nodes but have not yet been visited themselves. The known subgraph includes all edges of $G$ that are connected to the visited nodes $N_{\mathrm{V}}$. Importantly, naively stepping towards a node that is a locally optimal choice without incorporating information about the journey to date can trap a traveler in a dead end. Developing an algorithm with knowledge of $G_{\mathrm{K}}$ enables a traveler to navigate out of dead ends.  

\begin{figure}
\begin{centering}
\includegraphics{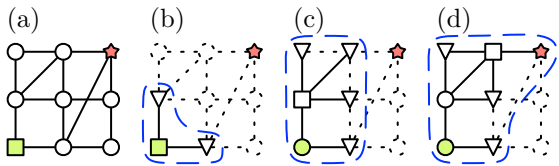}
\end{centering}
\caption{[Color online] (a) A lattice with the origin node represented by a square and the target node represented by a star. (b) The data available to a decentralized algorithm before the first step. We show a network traveler's current node as a square and the frontier nodes as triangles. We enclose the known subgraph $G_{\mathrm{K}}$ with a dashed contour.  Panel (c) shows the available data before the second step, and panel (d) shows the available data before the third step.
}
\label{fig:algorithm}
\end{figure}

In this local approach, we build on Fan et al.'s algorithm \cite{Fan2005Arriving, Fan2006Optimal} and apply it to $G_{\mathrm{K}}$ by changing the initial conditions of the sequences $\{v_{i}^{s}\left(t\right)\}$ and $\{w_{i}^{s}\left(t\right)\}$ for frontier nodes using the estimation function:
\[
	v_{j}^{0}\left(t\right)=w_{j}^{0}\left(t\right)=f\left(j,r;t\right)\,, \qquad \forall\, j\in N_{\mathrm{F}}\,.
\]
We initialize nodes that have been visited in the same manner as before (so $w_j^s \geq v_j^s$ is satisfied for all frontier nodes $j \in N_{\rm F}$). We iterate the two sets of sequences until they converge to within a chosen tolerance $\epsilon$. The traveler subsequently moves to a successor node that is identified by one of the criteria. We then update $G_{\mathrm{K}}$ and reduce the remaining budget by the weight incurred by making the step. We repeat this process until the traveler reaches the target or the budget is exhausted. Figure \ref{fig:algorithm} illustrates this routing process on a lattice.

We require $w_{i}^{k}\left(t\right)$ to provide upper bounds for the CDFs. In centralized routing, such an upper bound is equal to $1$ because the target is part of the network under consideration. In our decentralized situation, the upper bound for the CDF of a node in $G_{\mathrm{K}}$ is 
\[
	f_{\max}^r\left(t\right)=\max_{j\in N_{\mathrm{F}}}\left[f\left(r,j;t\right)\right]\,.
\]
Changing the initial conditions of visited nodes to
\[
	w_{i}^{0}\left(t\right)=f_{\max}\left(t\right)\,,\qquad\forall\, i\in N_{\mathrm{V}}
\]
accelerates the convergence of the two sets of sequences because it reduces the initial differences between them. In Appendix \ref{app:pseudo}, we give pseudocode for our decentralized algorithm.


\section{Simulations on a Small-World Network}
\label{sec:simulation} 

We test our algorithm on a variant of Kleinberg's small-world network \cite{Kleinberg2000The-small-world}.  We start with a $10\times10$ square lattice with undirected edges between neighboring nodes in the grid; we also assign an undirected shortcut edge from each node to exactly one other node.  For each shortcut, we determine the destination node using independent random trials so that the probability of such a long-range edge is proportional to $1/D_{ij}^2$, where $D_{ij}$ is the lattice distance between $i$ and $j$. (In determining shortcuts, we discard duplicate edges.) 

Noland et al. \cite{Noland2002Travel} (and references therein) investigated traffic-flow data sets and reported that travel times have log-normal distributions. For the purpose of numerical simulations, we let the random edge weights be distributed log-normally with PDF
\[
	p_{\ln}\left(\mu, \sigma; t\right) = \frac{1}{x\sqrt{2\pi\sigma^2}} \exp\left[-\frac{1}{2}\left(\frac{\ln x - \mu}{\sigma}\right)^2\right]\,.
\]
We assign a random variable to each edge by choosing the parameters $\mu$ and $\sigma$ uniformly at random from the interval $\left[0.5,1.5\right]$.

We consider Fan et al.'s centralized algorithm and our new decentralized algorithm with the joint criterion that we described previously. Note that the joint criterion reduces to Fan et al.'s criterion in the limit $\theta\rightarrow 1$. We let $d_{ij}$ be the Euclidean distance between nodes $i$ and $j$, and we approximate the network distance by $g_{ij} \approx h\left(d_{ij}\right) \equiv d_{ij}$. 

We investigate two different choices for the characteristic edge length $\lambda$ and the characteristic distribution $\bar{p}\left(t\right)$. First, we let $\lambda$ be the mean length of all edges in a network. Without further knowledge about a network, we assume that the weight $\bar{t}$ incurred by making a step towards the target node is chosen uniformly at random from the weights associated with the edges. The PDF of $\bar{t}$ is the mixture distribution \cite{Fruhwirth-Schnatter2006Finite}
\[
	\bar{p}\left(t\right)=\frac{1}{m}\sum_{\left(i,j\right)\in E}p_{ij}\left(t\right)\,.
\]
We call this method \emph{global estimation} (GE), because we make indirect use of global knowledge by using $\lambda$ and $\bar{p}\left(t\right)$ as a characteristic length scale and PDF, respectively.

Second, we restrict the sample from which we calculate $\lambda$ and $\bar{p}\left(t\right)$ to edges that are connected to visited nodes.  In this way, we only use local information.  In particular, we calculate
\begin{align*}
	E_{\mathrm{V}} &= \left\{(i,j) \in E \middle| i \in N_{\mathrm{V}}\right\}\,,\\
	\lambda        &= \frac{1}{\left|E_{\mathrm{V}}\right|}\sum_{\left(i,j\right)\in E_{\mathrm{V}}} d_{ij}\,,\\
	\bar{p}\left(t\right) &= \frac{1}{\left|E_{\mathrm{V}}\right|}\sum_{\left(i,j\right)\in E_{\mathrm{V}}} p_{ij}\left(t\right).\\
\end{align*}
We call this method \emph{local estimation} (LE).

\begin{figure}
\begin{centering}
\includegraphics{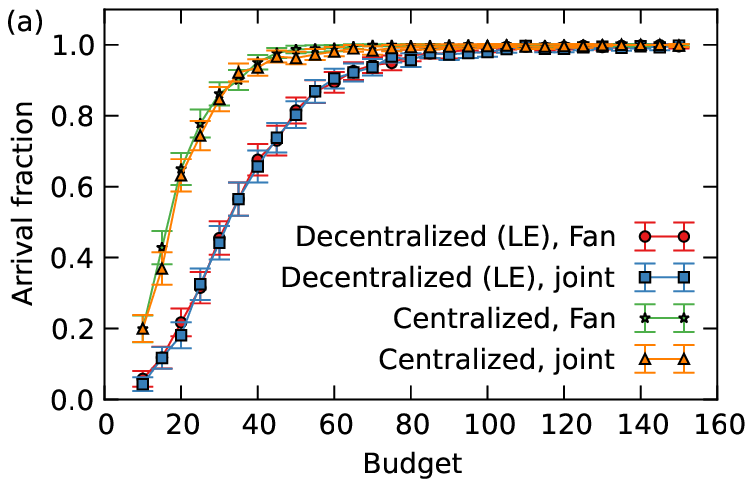}
\includegraphics{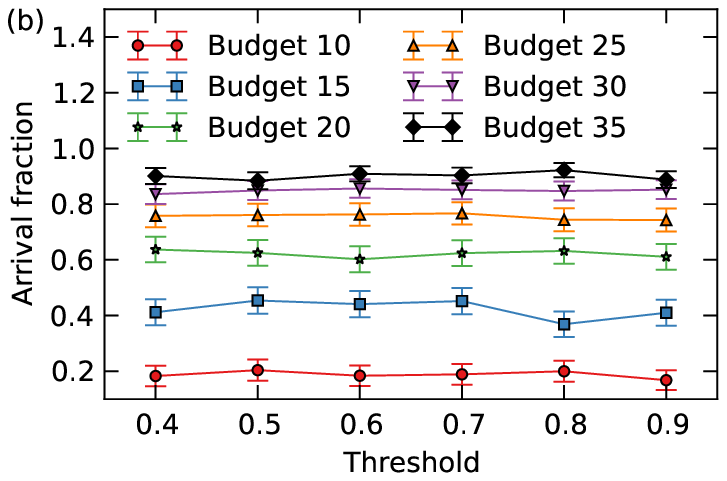}
\end{centering}
\caption{[Color online] (a) Fraction of routing attempts that successfully reach the target node in (a variant of) the Kleinberg network as a function of budget for a CDF threshold of $\theta = 0.8$ and a tolerance of $\epsilon=10^{-3}$. (We do not show the results of GE because they agree with those of LE.) (b) Arrival fractions obtained by Fan et al.'s centralized algorithm as a function of CDF threshold for a tolerance of $\epsilon=10^{-3}$ and several different budgets. The error bars in the two panels correspond to three standard deviations from the mean. }
\label{fig:lattice-arrival}
\end{figure}

Suppose that the origin of a routing process is $\left(2,2\right)$ and that the target is $\left(9,9\right)$, where $\left(x,y\right)$ designates a node using its lattice coordinates. We run each test $10^{3}$ times for several budgets [see Fig.~\ref{fig:lattice-arrival}(a)], several CDF thresholds [see Fig.~\ref{fig:lattice-arrival}(b)], and a tolerance of $\epsilon=10^{-3}$. (An error in arrival probability smaller than a tenth of a percent will not affect real travelers.)

As shown in Fig.~\ref{fig:lattice-arrival}(a), the arrival fraction---i.e., the fraction of routing attempts that reach the target node within a given budget---increases with increasing budget. Centralized algorithms know the entire network topology and can thus make better decisions, so they have larger arrival fractions. The LE algorithms have the same arrival fraction as the GE algorithms. Thus, it is sufficient to sample the network locally and global knowledge is not required as long as the network is sufficiently homogeneous (i.e., if a sample of the network provides representative estimates of $\lambda$ and $\bar{p}$). In some cases, we note that some global characteristics of a network might even be known \emph{a priori}, and such information can be used to inform sampling strategies. 

As shown in Fig.~\ref{fig:lattice-arrival}(b), the arrival fraction of Fan et al.'s centralized algorithm using the joint criterion is almost independent of the threshold. Because the local neighbors of each node are located in the cardinal directions, the arrival CDFs are sufficiently different for CDF maximization and travel time minimization to agree. We obtain the same results using our decentralized algorithm. Travelers thus choose the same successor node irrespective of the threshold $\theta$; this results in the same arrival fraction.

\begin{figure}
\begin{centering}
\includegraphics{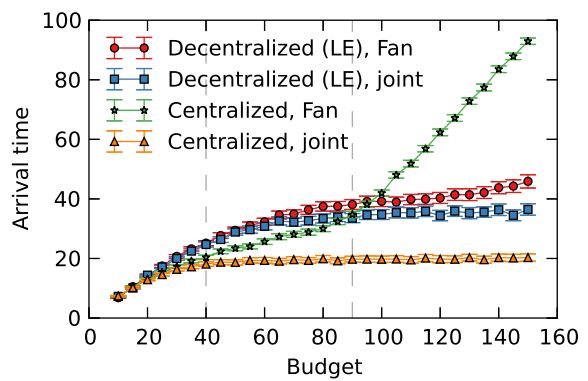}
\end{centering}
\caption{[Color online] Mean travel time of successful routing attempts on (a variant of) the Kleinberg network as a function of budget. We separate small, intermediate, and large budgets using dashed vertical lines.  (We do not show the results of GE because they agree with those of LE.) The error bars correspond to three standard deviations from the mean.}
\label{fig:lattice-time}
\end{figure}

Because centralized algorithms are aware of all shortcuts in a network, they have smaller mean travel times than decentralized ones (see Fig.~\ref{fig:lattice-time}). For small budgets ($\tau\lesssim 40$), the travel times of all algorithms increase with increasing budget. The algorithms choose neighboring nodes to maximize the CDF, which results in longer travel times because it is advantageous to exhaust the budget. For budgets $\tau$ that satisfy $40\lesssim\tau\lesssim 90$, the travel times using Fan et al.'s criterion and our joint criterion start to differ. The joint criterion starts to minimize travel times in this regime until they approach a steady value. Fan et al.'s criterion, however, continues to maximize CDFs such that travel times grow with increasing budget. For larger budgets ($\tau\gtrsim 90$), Fan et al.'s criterion is unable to distinguish between the CDFs of neighboring nodes, and algorithms using this criterion enter an unguided phase (i.e., one can construe a traveler to be ``lost"). The algorithm steps to neighbor nodes seemingly at random until the budget decreases sufficiently for Fan et al.'s criterion to discriminate among CDFs. As illustrated in Fig.~\ref{fig:lattice-time}, the travel time using Fan et al.'s criterion increases linearly with the budget in this regime.

See Ref.~\cite{Nie2009Shortest} for a thorough comparison of the arrival fraction for algorithms that treat edge weights as stochastic variables versus algorithms that minimize expected travel time.

\section{Simulations on the Chicago Sketch Network}\label{chicago}

\begin{figure}
	\includegraphics[width = .8\columnwidth]{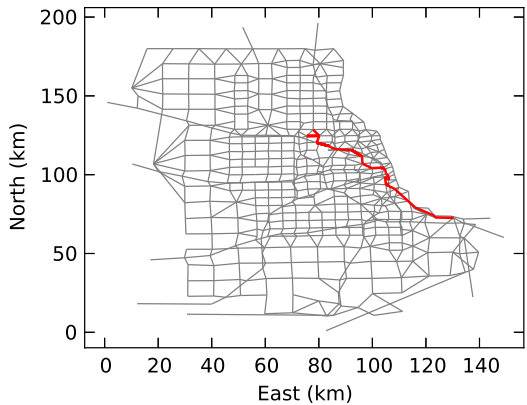}
\caption{[Color online] The Chicago sketch network, which has $n=542$ nodes and $m=1084$ edges. We show a path in red.}
\label{fig:chicago}
\end{figure}

We also tested our algorithm on the Chicago sketch network (CSN), which representes an aggregated version of the Chicago metropolitan road network that was developed and provided by the Chicago Area Transportation Study \cite{tntp, eash1983equilibrium}.  The CSN network, which we show in Fig.~\ref{fig:chicago}, has $n=542$ nodes and $m=1084$ edges.

\begin{figure}[b]
	\includegraphics[width = .8\columnwidth]{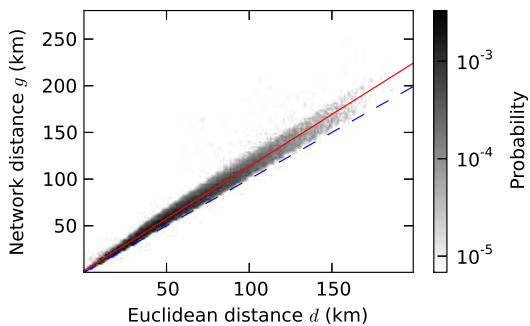}
	\caption{[Color online] Joint PDF for the Euclidean distance $d_{ij}$ and network distance $g_{ij}$ on the Chicago sketch network. The Pearson correlation coefficient between the two distance measures is $\rho \approx 0.985$, which justifies the linear fit (bright red line). The blue dashed line corresponds to the lower bound on the network distance (i.e., $g=d$).}
\label{fig:chicago-metric}
\end{figure}

In Section \ref{sec:estimation}, we claimed that there exists a function $h$ such that the network distance between two nodes $i$ and $j$ is well-approximated by $h\left(d_{ij}\right)$. Our investigation of the CSN allow us to examine this claim more closely. In Fig.~\ref{fig:chicago-metric}, we show the joint PDF for the Euclidean distance $d_{ij}$ and network distance $g_{ij}$ for the CSN. To estimate the PDF, we calculate the Euclidean and network distance for all $\frac{1}{2}n\left(n-1\right)=146611$ distinct pairs of nodes and bin the data on a $200\times 200$ grid.

The Euclidean distance
\[
	d_{ij}=\sqrt{\left(x_i-x_j\right)^2 + \left(y_i-y_j\right)^2}
\]
is strongly correlated with the network distance. Based on a linear bootstrap fit \cite{Efron1993An-introduction}, the best choice for $h\left(d_{ij}\right)\approx g_{ij}$ is
\[
	h\left(d_{ij}\right) \approx 1.67(2)\text{ km} + 1.1176(4) \times d_{ij},
\]
where $d_{ij}$ has units of km. [Recall that ``1.67(2)" means that the error bars place the value between 1.65 and 1.69.] The slope of the linear fit is larger than 1 because the Euclidean distance between each pair of nodes provides a lower bound for the network distance between those two nodes.

We obtain similar results when using the lattice distance 
\begin{equation}
	d_{ij}=\lvert x_i-x_j\rvert + \lvert y_i-y_j\rvert
	\label{eq:lattice}
\end{equation}
between nodes $i$ and $j$. In this case, the slope of the linear fit is smaller than 1 because the lattice distance provides an approximate upper bound for the network distance. Complicated paths, such as zigzag paths, can of course violate this approximate bound.

The mean edge length of the network is $\lambda \approx 5.77\text{ km}$. Note that the mean edge length exceeds the typical length of roads in metropolitan areas because the CSN is aggregated: there is not a one-to-one correspondence between nodes and junctions.  We choose the origin and target nodes uniformly at random such that their Euclidean distance lies in the interval $[40, 50]\text{ km}$. We consider the same four tests as in Section \ref{sec:simulation} and run each test $10^3$ times for several budgets, several CDF thresholds, and a numerical tolerance of $\epsilon=10^{-3}$. 

\begin{figure}
\begin{centering}
\includegraphics{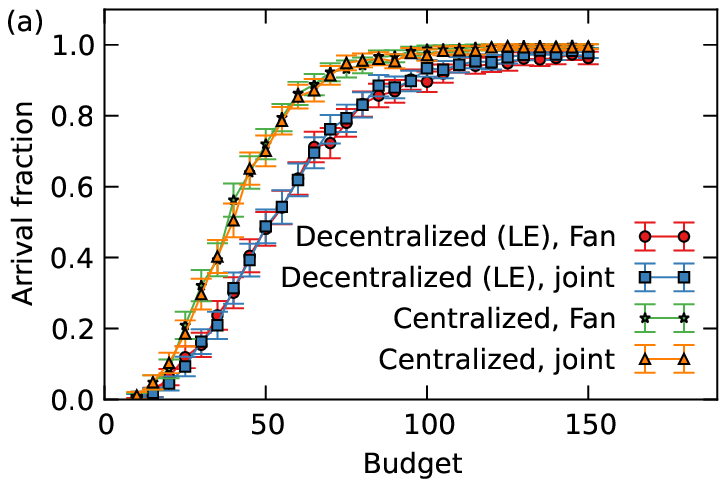} 
\hspace{.5 in}
\includegraphics{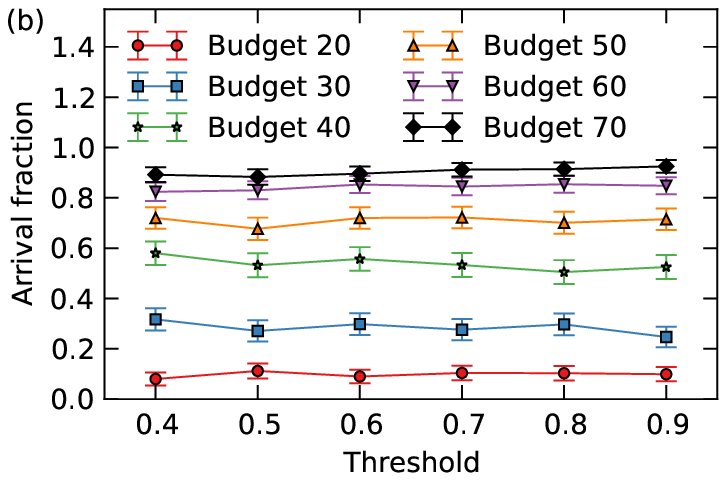}
\end{centering}
\caption{[Color online] (a) Fraction of routing attempts that successfully reach the target on the CSN as a function of budget for a CDF threshold of $\theta = 0.8$ and a tolerance of $\epsilon=10^{-3}$. 
(We do not show the results of GE because they agree with those of LE.) (b) Arrival fractions obtained by Fan et al.'s centralized algorithm as a function of CDF threshold for a tolerance of $\epsilon=10^{-3}$ and several different budgets. The error bars in the two panels correspond to three standard deviations from the mean.}
\label{fig:chicago-arrival}
\end{figure}

\begin{figure}
\begin{centering}
\includegraphics[width = .8\columnwidth]{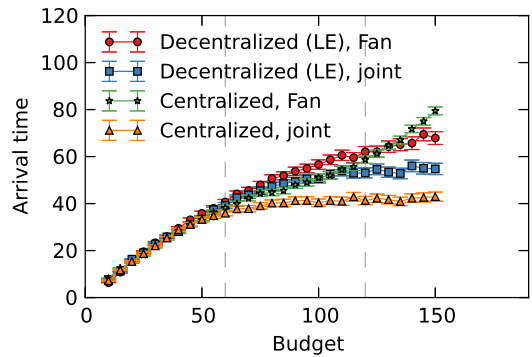}
\end{centering}
\caption{[Color online] Mean travel time of successful routing attempts on the CSN as a function of budget. We separate small, intermediate, and large budgets using dashed vertical lines. (We do not show the results of GE because they agree with those of LE.) The error bars correspond to three standard deviations from the mean.}
\label{fig:chicago-time}
\end{figure}

In Fig.~\ref{fig:chicago-arrival}(a), we illustrate the arrival fractions as a function of budget.  We observe the same qualitative behavior as on the (variant of the) Kleinberg small-world network. As with the Kleinberg network, the arrival fraction of algorithms that use the joint criterion depends very little on the CDF threshold $\theta$ [see Fig.~\ref{fig:chicago-arrival}(b)]. In Fig.~\ref{fig:chicago-time}, we show that the travel times of centralized algorithms are smaller than those of decentralized algorithms (because the former know all shortcuts in the network).

\section{Additional Remarks} \label{add}

\begin{figure}
\begin{centering}
\includegraphics[width = .8\columnwidth]{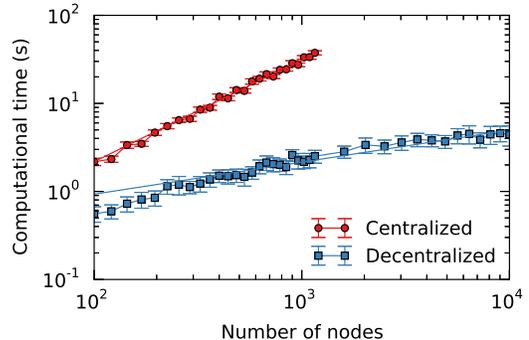}
\end{centering}
\caption{[Color online] Mean computational times for Fan et al.'s centralized algorithm and our decentralized algorithm as a function of the number of nodes of a Kleinberg small-world network. The fits represent power-law scalings with exponents of $1.16$ (Fan et al.) and $0.36$ (decentralized). The error bars correspond to three standard deviations from the mean.  We determined the values of these exponents using mean-square regression, and we note that it does not matter whether these curves follow precise power laws.
\label{fig:computational-time}}
\end{figure}

To investigate the computational time of Fan et al.'s algorithm and our decentralized algorithm, we perform 100 simulations for several sizes of a Kleinberg small-world network. Fan et al.'s algorithm needs to consider the entire network to construct a routing table (see Section \ref{adapt}). Hence, its computational time increases approximately linearly with the number of nodes (see Fig.~\ref{fig:computational-time}). However, our decentralized algorithm only considers nodes that are near nodes it has already visited, so its computational time increases sublinearly with the number of nodes.  This sublinear scaling illustrates that decentralized routing is possible for networks with stochastic edge weights because travelers do not explore networks uniformly.

Fan et al.'s algorithm is more appropriate if a network is static and the PDFs do not change, because one can use the same routing table for all travelers on a network who wish to reach the same destination node. Our decentralized algorithm is more appropriate if edges appear and/or disappear or, more generally, if the PDFs change during the routing process.  Thus, our decentralized algorithm is a more appropriate match for applications to traveling in real life.

Another interesting problem is routing on a network whose topology is known but whose edge-weight distributions are unknown. Without any knowledge about the PDFs, we assume that all edge weights are independently and identically distributed with PDF $\hat{p}(t)$. The arrival CDF along some path $\ell$ depends only on the number of steps $k$ and is given by 
\begin{equation*}
	U_k\left(t\right)=\displaystyle \int_{0}^{t}dt'\,\left(\mathop{\ast}\limits _{k'=1}^{k}\hat{p}\right)\left(t'\right)\,.
\end{equation*} 
Fan et al. \cite{Fan2006Optimal} showed that $U_k\left(t\right)$ is a non-decreasing series in $k$.

Because the criteria in Section \ref{crit} favor larger values of $U_k\left(t\right)$, paths with the smallest number of steps are optimal. Hence, the problem reduces to a DSSP unless some information about the edge weights is available.


\section{Conclusions} \label{conc}

We have examined decentralized routing on networks with stochastic edges weights. Our contributions are twofold. First, we have introduced a new criterion to discriminate among the CDFs of paths. Our criterion circumvents the limitations of the criteria proposed by Fan et al. and Frank, but it retains the desirable properties of both because it minimizes travel times without sacrificing reliability. It also provides a better caricature of the behavior of real travelers \cite{Small1982The-Scheduling}.  Second, we have developed a decentralized routing algorithm that is applicable to networks with stochastic edge weights. Our algorithm employs a CDF estimation function that captures a notion of proximity in space and guides network travelers without the need to incorporate global knowledge about a network. Our simulation results demonstrate that decentralized routing on networks with stochastic edge weights is viable.  

Our approach appears to be very promising. Investigating both its limitations and the situations in which it is most successful are important topics for future research. In particular, it is important to examine the effects of inhomogeneities and different classes of PDFs on routing performance. Possible improvements of our algorithm include the development of more sophisticated choices of estimation functions that incorporate edge lengths, edge weights, and their correlations. We expect such work to be particularly interesting in studies of routing on temporal networks, in which the existence and other properties of edges are time-dependent.


\section*{Acknowlegements} 

We thank Aaron Clauset, Hillel Bar-Gera, Jon Kleinberg, Sang Hoon Lee, Peter Mucha, and Jie Sun for useful discussions and the referees for their helpful comments. MAP was supported by the James S. McDonnell Foundation (research award number 220020177), the EPSRC (EP/J001759/1), and the FET-Proactive project PLEXMATH (FP7-ICT-2011-8; grant number 317614) funded by the European Commission. He also thanks some mathematicians from University of Bath who came to Oxford (because of a convenient workshop) and returned his umbrella. This paper presents research results of the Belgian Network DYSCO, which were funded by the IAP Programme and initiated by Beslpo.



\begin{thebibliography}{10}%
\makeatletter
\providecommand \@ifxundefined [1]{%
 \ifx #1\undefined \expandafter \@firstoftwo
 \else \expandafter \@secondoftwo
\fi
}%
\providecommand \@ifnum [1]{%
 \ifnum #1\expandafter \@firstoftwo
 \else \expandafter \@secondoftwo
\fi
}%
\providecommand \enquote [1]{``#1''}%
\providecommand \bibnamefont  [1]{#1}%
\providecommand \bibfnamefont [1]{#1}%
\providecommand \citenamefont [1]{#1}%
\providecommand\href[0]{\@sanitize\@href}%
\providecommand\@href[1]{\endgroup\@@startlink{#1}\endgroup\@@href}%
\providecommand\@@href[1]{#1\@@endlink}%
\providecommand \@sanitize [0]{\begingroup\catcode`\&12\catcode`\#12\relax}%
\@ifxundefined \pdfoutput {\@firstoftwo}{%
 \@ifnum{\z@=\pdfoutput}{\@firstoftwo}{\@secondoftwo}%
}{%
 \providecommand\@@startlink[1]{\leavevmode}%
 \providecommand\@@endlink[0]{}%
}{%
 \providecommand\@@startlink[1]{%
  \leavevmode
  \pdfstartlink
   attr{/Border[0 0 1 ]/H/I/C[0 1 1]}%
   user{/Subtype/Link/A<</Type/Action/S/URI/URI(#1)>>}%
  \relax
 }%
 \providecommand\@@endlink[0]{\pdfendlink}%
}%
\providecommand \url  [0]{\begingroup\@sanitize \@url }%
\providecommand \@url [1]{\endgroup\@href {#1}{\urlprefix}}%
\providecommand \urlprefix [0]{URL }%
\providecommand \Eprint[0]{\href }%
\@ifxundefined \urlstyle {%
  \providecommand \doi [1]{doi:\discretionary{}{}{}#1}%
}{%
  \providecommand \doi [0]{doi:\discretionary{}{}{}\begingroup
  \urlstyle{rm}\Url }%
}%
\providecommand \doibase [0]{http://dx.doi.org/}%
\providecommand \Doi[1]{\href{\doibase#1}}%
\providecommand \bibAnnote [3]{%
  \BibitemShut{#1}%
  \begin{quotation}\noindent
    \textsc{Key:}\ #2\\\textsc{Annotation:}\ #3%
  \end{quotation}%
}%
\providecommand \bibAnnoteFile [2]{%
  \IfFileExists{#2}{\bibAnnote {#1} {#2} {\input{#2}}}{}%
}%
\providecommand \typeout [0]{\immediate \write \m@ne }%
\providecommand \selectlanguage [0]{\@gobble}%
\providecommand \bibinfo [0]{\@secondoftwo}%
\providecommand \bibfield [0]{\@secondoftwo}%
\providecommand \translation [1]{[#1]}%
\providecommand \BibitemOpen[0]{}%
\providecommand \bibitemStop [0]{}%
\providecommand \bibitemNoStop [0]{.\EOS\space}%
\providecommand \EOS [0]{\spacefactor3000\relax}%
\providecommand \BibitemShut [1]{\csname bibitem#1\endcsname}%
\bibitem{Clauset2003How-Do-Networks}%
  \BibitemOpen
  \bibfield{author}{%
  \bibinfo {author} {\bibnamefont{{A. {Clauset} and C. {Moore} }}}}%
   (\bibinfo {year} {2003}),\
  \Eprint{http://arxiv.org/abs/arXiv:cond-mat/0309415}{arXiv:cond-mat/0309415}%
  \bibAnnoteFile{NoStop}{Clauset2003How-Do-Networks}%
\bibitem{Boguna2009Navigability}%
  \BibitemOpen
  \bibfield{author}{%
  \bibinfo {author} {\bibfnamefont{M.}~\bibnamefont{{Bogu{\~n}{\'a}}}},
  \bibinfo {author} {\bibfnamefont{D.}~\bibnamefont{{Krioukov}}},\ and\
  \bibinfo {author} {\bibfnamefont{K.~C.}\ \bibnamefont{{Claffy}}},\ }%
  \bibfield{journal}{%
  \bibinfo {journal} {Nature Physics}\ }%
  \textbf{\bibinfo {volume} {5}},\ \bibinfo {pages} {74} (\bibinfo {year}
  {2009})%
  \bibAnnoteFile{NoStop}{Boguna2009Navigability}%
\bibitem{Erola2011Structural}%
  \BibitemOpen
  \bibfield{author}{%
  \bibinfo {author} {\bibfnamefont{P.}~\bibnamefont{Erola}}, \bibinfo {author}
  {\bibfnamefont{S.}~\bibnamefont{G{\'o}mez}},\ and\ \bibinfo {author}
  {\bibfnamefont{A.}~\bibnamefont{Arenas}},\ }%
  \bibfield{journal}{%
  \bibinfo {journal} {Int. J. Complex Systems in Science}\ }%
  \textbf{\bibinfo {volume} {1}},\ \bibinfo {pages} {37} (\bibinfo {year}
  {2011})%
  \bibAnnoteFile{NoStop}{Erola2011Structural}%
\bibitem{Newman2010Networks}%
  \BibitemOpen
  \bibfield{author}{%
  \bibinfo {author} {\bibfnamefont{M.~E.~J.}\ \bibnamefont{Newman}},\ }%
  \emph{\bibinfo {title} {Networks: An Introduction}}\ (\bibinfo {publisher}
  {Oxford University Press},\ \bibinfo {year} {2010})%
  \bibAnnoteFile{NoStop}{Newman2010Networks}%
\bibitem{Dijkstra1959A-note}%
  \BibitemOpen
  \bibfield{author}{%
  \bibinfo {author} {\bibfnamefont{E.}~\bibnamefont{Dijkstra}},\ }%
  \bibfield{journal}{%
  \bibinfo {journal} {Numerische Mathematik}\ }%
  \textbf{\bibinfo {volume} {1}},\ \bibinfo {pages} {269} (\bibinfo {year}
  {1959})%
  \bibAnnoteFile{NoStop}{Dijkstra1959A-note}%
\bibitem{Milgram1967The-small}%
  \BibitemOpen
  \bibfield{author}{%
  \bibinfo {author} {\bibfnamefont{S.}~\bibnamefont{Milgram}},\ }%
  \bibfield{journal}{%
  \bibinfo {journal} {Psychology Today}\ }%
  \textbf{\bibinfo {volume} {2}},\ \bibinfo {pages} {60} (\bibinfo {year}
  {1967})%
  \bibAnnoteFile{NoStop}{Milgram1967The-small}%
\bibitem{Berman1998On-line}%
  \BibitemOpen
  \bibfield{author}{%
  \bibinfo {author} {\bibfnamefont{P.}~\bibnamefont{Berman}},\ }%
  in\ \emph{\bibinfo {booktitle} {Online Algorithms: The State of the Art}},\ \bibinfo {series and number} {Lecture Notes in Computer Science, Volume 1442 (Editors: A. Fiat and G.~J. Woeginger)}\
  (\bibinfo {publisher} {Springer Verlag},\ \bibinfo {address} {Berlin, Germany},\
  \bibinfo {year} {1998})\ pp.\ \bibinfo {pages} {232--241}%
  \bibAnnoteFile{NoStop}{Berman1998On-line}%
\bibitem{Peleg1989A-trade-off}%
  \BibitemOpen
  \bibfield{author}{%
  \bibinfo {author} {\bibfnamefont{D.}~\bibnamefont{Peleg}}\ and\ \bibinfo
  {author} {\bibfnamefont{E.}~\bibnamefont{Upfal}},\ }%
  \bibfield{journal}{%
  \bibinfo {journal} {Journal of the ACM}\ }%
  \textbf{\bibinfo {volume} {36}},\ \bibinfo {pages} {510} (\bibinfo {year}
  {1989})%
  \bibAnnoteFile{NoStop}{Peleg1989A-trade-off}%
\bibitem{Krioukov2007On-compact}%
  \BibitemOpen
  \bibfield{author}{%
  \bibinfo {author} {\bibfnamefont{D.}~\bibnamefont{Krioukov}}, \bibinfo
  {author} {\bibfnamefont{K.~C.}\ \bibnamefont{Claffy}}, \bibinfo {author}
  {\bibfnamefont{K.}~\bibnamefont{Fall}},\ and\ \bibinfo {author}
  {\bibfnamefont{A.}~\bibnamefont{Brady}},\ }%
  \bibfield{journal}{%
  \bibinfo {journal} {SIGCOMM Comput. Commun. Rev.}\ }%
  \textbf{\bibinfo {volume} {37}},\ \bibinfo {pages} {41} (\bibinfo {year}
  {2007})%
  \bibAnnoteFile{NoStop}{Krioukov2007On-compact}%
\bibitem{Kleinberg2000Navigation}%
  \BibitemOpen
  \bibfield{author}{%
  \bibinfo {author} {\bibfnamefont{J.~M.}\ \bibnamefont{Kleinberg}}
  \emph{et~al.},\ }%
  \bibfield{journal}{%
  \bibinfo {journal} {Nature}\ }%
  \textbf{\bibinfo {volume} {406}},\ \bibinfo {pages} {845} (\bibinfo {year}
  {2000})%
  \bibAnnoteFile{NoStop}{Kleinberg2000Navigation}%
\bibitem{barth2011}%
  \BibitemOpen
  \bibfield{author}{%
  \bibinfo {author} {\bibfnamefont{M.}~\bibnamefont{Barth\'elemy}},\ }%
  \bibfield{journal}{%
  \bibinfo {journal} {Physics Reports}\ }%
  \textbf{\bibinfo {volume} {49}},\ \bibinfo {pages} {1} (\bibinfo {year}
  {2011})%
  \bibAnnoteFile{NoStop}{barth2011}%
\bibitem{Rosvall2005Navigating}%
  \BibitemOpen
  \bibfield{author}{%
  \bibinfo {author} {\bibfnamefont{M.}~\bibnamefont{{Rosvall}}}, \bibinfo
  {author} {\bibfnamefont{P.}~\bibnamefont{{Minnhagen}}},\ and\ \bibinfo
  {author} {\bibfnamefont{K.}~\bibnamefont{{Sneppen}}},\ }%
  \bibfield{journal}{%
  \bibinfo {journal} {\pre}\ }%
  \textbf{\bibinfo {volume} {71}},\ \bibinfo {eid} {066111} (\bibinfo {year}
  {2005})%
  \bibAnnoteFile{NoStop}{Rosvall2005Navigating}%
\bibitem{Liben-Nowell2005Geographic}%
  \BibitemOpen
  \bibfield{author}{%
  \bibinfo {author} {\bibfnamefont{D.}~\bibnamefont{Liben-Nowell}}, \bibinfo
  {author} {\bibfnamefont{J.}~\bibnamefont{Novak}}, \bibinfo {author}
  {\bibfnamefont{R.}~\bibnamefont{Kumar}}, \bibinfo {author}
  {\bibfnamefont{P.}~\bibnamefont{Raghavan}},\ and\ \bibinfo {author}
  {\bibfnamefont{A.}~\bibnamefont{Tomkins}},\ }%
  \bibfield{journal}{%
  \bibinfo {journal} {Proceedings of the National Academy of Sciences of the
  United States of America}\ }%
  \textbf{\bibinfo {volume} {102}},\ \bibinfo {pages} {11623} (\bibinfo {year}
  {2005})%
  \bibAnnoteFile{NoStop}{Liben-Nowell2005Geographic}%
\bibitem{rio10}%
  \BibitemOpen
  \bibfield{author}{%
  \bibinfo {author} {\bibnamefont{{J. {Sun} and D. {ben-Avraham} }}}}%
   (\bibinfo {year} {2010}),\
  \Eprint{http://arxiv.org/abs/1001.5196}{arXiv:1001.5196}%
  \bibAnnoteFile{NoStop}{rio10}%
\bibitem{Aldous2010Connected}%
  \BibitemOpen
  \bibfield{author}{%
  \bibinfo {author} {\bibnamefont{{D.~J. {Aldous} and J. {Shun}, }}}}%
 \bibinfo {journal} {Statistical Science }%
  \textbf{\bibinfo {volume} {25}},\ \bibinfo {pages} {275} (\bibinfo {year}
  {2010})%
  \bibAnnoteFile{NoStop}{Aldous2010Connected}
\bibitem{Lee2011Pathlength}%
  \BibitemOpen
  \bibfield{author}{%
  \bibinfo {author} {\bibfnamefont{S.~H.}\ \bibnamefont{Lee}}\ and\ \bibinfo
  {author} {\bibfnamefont{P.}~\bibnamefont{Holme}},\ }%
  \bibfield{journal}{%
  \bibinfo {journal} {Physica A}\ }%
  \textbf{\bibinfo {volume} {390}},\ \bibinfo {pages} {3996} (\bibinfo {year}
  {2011})%
  \bibAnnoteFile{NoStop}{Lee2011Pathlength}%
\bibitem{Lee2012Geometric}%
  \BibitemOpen
  \bibfield{author}{%
  \bibinfo {author} {\bibnamefont{{S.~H. {Lee} and P. {Holme}, }}}}%
  \bibfield{journal}{%
  \bibinfo {journal} {Physical Review E}\ }%
  \textbf{\bibinfo {volume} {86}},\ \bibinfo {pages} {067103} (\bibinfo {year}
  {2012})%
  \bibAnnoteFile{NoStop}{Lee2012Geometric}%
\bibitem{Noland2002Travel}%
  \BibitemOpen
  \bibfield{author}{%
  \bibinfo {author} {\bibfnamefont{R.~B.}\ \bibnamefont{Noland}}\ and\ \bibinfo
  {author} {\bibfnamefont{J.~W.}\ \bibnamefont{Polak}},\ }%
  \bibfield{journal}{%
  \bibinfo {journal} {Transport Reviews}\ }%
  \textbf{\bibinfo {volume} {22}},\ \bibinfo {pages} {39} (\bibinfo {year}
  {2002})%
  \bibAnnoteFile{NoStop}{Noland2002Travel}%
\bibitem{Fan2005Arriving}%
  \BibitemOpen
  \bibfield{author}{%
  \bibinfo {author} {\bibfnamefont{Y.~Y.}\ \bibnamefont{Fan}}, \bibinfo
  {author} {\bibfnamefont{R.~E.}\ \bibnamefont{Kalaba}},\ and\ \bibinfo
  {author} {\bibfnamefont{J.~E.}\ \bibnamefont{Moore}},\ }%
  \bibfield{journal}{%
  \bibinfo {journal} {Journal of Optimization Theory and Applications}\ }%
  \textbf{\bibinfo {volume} {127}},\ \bibinfo {pages} {497} (\bibinfo {year}
  {2005})%
  \bibAnnoteFile{NoStop}{Fan2005Arriving}%
\bibitem{Nie2009Shortest}%
  \BibitemOpen
  \bibfield{author}{%
  \bibinfo {author} {\bibfnamefont{Y.~M.}\ \bibnamefont{Nie}}\ and\ \bibinfo
  {author} {\bibfnamefont{X.}~\bibnamefont{Wu}},\ }%
  \bibfield{journal}{%
  \bibinfo {journal} {Transportation Research Part B: Methodological}\ }%
  \textbf{\bibinfo {volume} {43}},\ \bibinfo {pages} {597} (\bibinfo {year}
  {2009})%
  \bibAnnoteFile{NoStop}{Nie2009Shortest}%
\bibitem{Nie2006}%
  \BibitemOpen
  \bibinfo {author}{%
  {\bibfnamefont{Y.}\ \bibnamefont{Nie}}\ and\ \bibinfo
  {author} {\bibfnamefont{Y.}~\bibnamefont{Fan}},\ }%
  \bibfield{journal}{%
  \bibinfo {journal} {Transportation Research Record: Journal of the Transportation Research Board}\ }%
  \textbf{\bibinfo {volume} {1964}},\ \bibinfo {pages} {193} (\bibinfo {year}
  {2006})%
  \bibAnnoteFile{NoStop}{Nie2006}%
\bibitem{Frank1969Shortest}%
  \BibitemOpen
  \bibfield{author}{%
  \bibinfo {author} {\bibfnamefont{H.}~\bibnamefont{Frank}},\ }%
  \bibfield{journal}{%
  \bibinfo {journal} {Operations Research}\ }%
  \textbf{\bibinfo {volume} {17}},\ \bibinfo {pages} {583} (\bibinfo {year}
  {1969})%
  \bibAnnoteFile{NoStop}{Frank1969Shortest}%
\bibitem{Loui1983Optimal}%
  \BibitemOpen
  \bibfield{author}{%
  \bibinfo {author} {\bibfnamefont{R.~P.}\ \bibnamefont{Loui}},\ }%
  \bibfield{journal}{%
  \bibinfo {journal} {Commun. ACM}\ }%
  \textbf{\bibinfo {volume} {26}},\ \bibinfo {pages} {670} (\bibinfo {month}
  {September}\ \bibinfo {year} {1983})%
  \bibAnnoteFile{NoStop}{Loui1983Optimal}%
\bibitem{Eiger1985Path}%
  \BibitemOpen
  \bibfield{author}{%
  \bibinfo {author} {\bibfnamefont{A.}~\bibnamefont{Eiger}}, \bibinfo {author}
  {\bibfnamefont{P.~B.}\ \bibnamefont{Mirchandani}},\ and\ \bibinfo {author}
  {\bibfnamefont{H.}~\bibnamefont{Soroush}},\ }%
  \bibfield{journal}{%
  \bibinfo {journal} {Transportation Science}\ }%
  \textbf{\bibinfo {volume} {19}},\ \bibinfo {pages} {75} (\bibinfo {year}
  {1985})%
  \bibAnnoteFile{NoStop}{Eiger1985Path}%
\bibitem{Hoel1971Introduction}%
  \BibitemOpen
  \bibfield{author}{%
  \bibinfo {author} {\bibfnamefont{P.~G.}\ \bibnamefont{Hoel}}, \bibinfo
  {author} {\bibfnamefont{S.~C.}\ \bibnamefont{Port}},\ and\ \bibinfo {author}
  {\bibfnamefont{C.~J.}\ \bibnamefont{Stone}},\ }%
  \emph{\bibinfo {title} {Introduction to Probability Theory}},\ Houghton
  Mifflin series in statistics\ (\bibinfo {publisher} {Houghton Mifflin},\
  \bibinfo {year} {1971})%
  \bibAnnoteFile{NoStop}{Hoel1971Introduction}%
\bibitem{Fan2006Optimal}%
  \BibitemOpen
  \bibfield{author}{%
  \bibinfo {author} {\bibfnamefont{Y.}~\bibnamefont{Fan}}\ and\ \bibinfo
  {author} {\bibfnamefont{Y.}~\bibnamefont{Nie}},\ }%
  \bibfield{journal}{%
  \bibinfo {journal} {Networks and Spatial Economics}\ }%
  \textbf{\bibinfo {volume} {6}},\ \bibinfo {pages} {333} (\bibinfo {year}
  {2006})%
  \bibAnnoteFile{NoStop}{Fan2006Optimal}%
\bibitem{Fruhwirth-Schnatter2006Finite}%
  \BibitemOpen
  \bibfield{author}{%
  \bibinfo {author} {\bibfnamefont{S.}~\bibnamefont{Fr{\"u}hwirth-Schnatter}},\
  }%
  \emph{\bibinfo {title} {Finite Mixture and Markov Switching Models}}\
  (\bibinfo {publisher} {Springer Verlag},\ \bibinfo {year} {2006})%
  \bibAnnoteFile{NoStop}{Fruhwirth-Schnatter2006Finite}%
\bibitem{Kleinberg2000The-small-world}%
  \BibitemOpen
  \bibfield{author}{%
  \bibinfo {author} {\bibfnamefont{J.}~\bibnamefont{Kleinberg}},\ }%
  in\ \emph{\bibinfo {booktitle} {Proceedings of the Thirty-Second Annual ACM
  Symposium on Theory of Computing}},\ \bibinfo {series and number} {STOC '00}\
  (\bibinfo {publisher} {ACM},\ \bibinfo {address} {New York, NY, USA},\
  \bibinfo {year} {2000})\ pp.\ \bibinfo {pages} {163--170}%
  \bibAnnoteFile{NoStop}{Kleinberg2000The-small-world}%
\bibitem{Small1982The-Scheduling}%
  \BibitemOpen
  \bibfield{author}{%
  \bibinfo {author} {\bibfnamefont{K.~A.}\ \bibnamefont{Small}},\ }%
  \bibfield{journal}{%
  \bibinfo {journal} {The American Economic Review}\ }%
  \textbf{\bibinfo {volume} {72}},\ \bibinfo {pages} {pp. 467} (\bibinfo {year}
  {1982})%
  \bibAnnoteFile{NoStop}{Small1982The-Scheduling}%
\bibitem{tntp}%
  \BibitemOpen
  \bibfield{author}{%
  \bibinfo {author} {\bibfnamefont{H.}~\bibnamefont{Bar-Gera}},\ }%
  \enquote{\bibinfo {title} {Transportation Network Test Problems},}\ \bibinfo
  {howpublished} {\url{http://www.bgu.ac.il/~bargera/tntp/ } }%
  (\bibinfo {year}
  {2001--2013})%
  \bibAnnoteFile{NoStop}{tntp}%
  
\bibitem{Efron1993An-introduction}%
  \BibitemOpen
  \bibfield{author}{%
  \bibinfo {author} {\bibfnamefont{B.}~\bibnamefont{Efron}}\ and\ \bibinfo
  {author} {\bibfnamefont{R.}~\bibnamefont{Tibshirani}},\ }%
  \emph{\bibinfo {title} {An Introduction to the Bootstrap}},\ Monographs on
  Statistics and Applied Probability\ (\bibinfo {publisher} {Chapman \& Hall},\
  \bibinfo {year} {1993})%
  \bibAnnoteFile{NoStop}{Efron1993An-introduction}%
  
\bibitem{eash1983equilibrium}
  \BibitemOpen
  \bibfield{author}{
  \bibinfo{author}{\bibfnamefont{R.~W.}~\bibnamefont{Eash}}, 
  \bibinfo{author}{\bibfnamefont{K.~S.}~\bibnamefont{Chon}} and
  \bibinfo{author}{\bibfnamefont{D.~E.}~\bibnamefont{Boyce}}, }
  \bibinfo{journal}{Transportation Research Record}
  \textbf{994},
  \bibinfo{pages}{30} (1983)
  \bibAnnoteFile{NoStop}{eash1983equilibrium}
  
\end{thebibliography}

%


\newpage

\newpage

\appendix


\section{Pseudocode for Our Decentralized Algorithm} \label{app:pseudo}

In Algorithm \ref{alg:algorithm}, we give pseudocode for our decentralized routing algorithm for networks with stochastic edge weights.

\begin{algorithm*}[h]
\begin{algorithmic}[5]
	\Function{DecentralizedRouting}{$G$, origin, target, $\tau,\,\theta$, \textsc{Criterion}}
		\State $\left(\text{current, traveltime}\right) \gets \left(\mathrm{origin}, 0\right)$
		\State steps $\gets\left\{\left(\text{current, traveltime}\right)\right\}$
		\State $N_{\mathrm{V}}\gets\left\{\text{current}\right\}$
		\While{traveltime $\leq\tau$ \textbf{and} current $\neq$ target}
			\State $N_{\mathrm{F}}\gets i\qquad\forall\left\{\left(i,j\right)\in E: j \in N_{\mathrm{V}} \text{ \textbf{and} } i \notin N_{\mathrm{V}}\right\}$
			\Comment{Obtain frontier nodes.}
			\State $v_i^0\left(t\right)\gets w_i^0\left(t\right)\gets f\left(i,\text{target};t\right)\qquad\forall\, i \in N_{\mathrm{F}}$
			\Comment{Initialize frontier nodes.}
			\State $f_{\max}^{\mathrm{target}}\left(t\right) = \max_{i\in N_{\mathrm{F}}}\left[f\left(i,\text{target};t\right)\right]$
			\Comment{Obtain upper bound.}
			\State $v_j^0\left(t\right)\gets 0\qquad\forall\, j\in N_{\mathrm{V}}$
			\Comment{Initialize visited nodes.}
			\State $w_j^0\left(t\right)\gets f_{\max}^{\mathrm{target}}\left(t\right)\qquad\forall\, j \in N_{\mathrm{V}}$
			\State unstable $\gets N_{\mathrm{V}}$
			\State $s\gets 0$
			\While{current $\in$ unstable}
				\State $v_i^{s+1}\left(t\right)=\max_{j\in J_i}\left[\int_0^t p_{ij}\left(t'\right)v_j^s\left(t-t'\right)\,dt'\right]\qquad\forall\, i\in$ unstable
				\Comment{Update the sequences.}
				\State $w_i^{s+1}\left(t\right)=\max_{j\in J_i}\left[\int_0^t p_{ij}\left(t'\right)w_j^s\left(t-t'\right)\,dt'\right]\qquad\forall\, i\in$ unstable
				\State $s\gets s+1$
				\For{$i\in$ unstable}
					\If{$\lvert v_i^s\left(t\right)-w_i^s\left(t\right)\rvert<\epsilon\qquad\forall\, t$}
					\Comment{Check for convergence.}
						\State \textbf{remove} $i$ \textbf{from} unstable
					\EndIf
				\EndFor
			\EndWhile
			\State $q_{\text{current}}\left(t\right)=\mathop{\arg\max}_{j\in J_{\text{current}}}\left[\int_0^t p_{\text{current }j}\left(t'\right)v_j^s\left(t-t'\right)\,dt'\right]$
			\State successor $\gets$ \Call{Criterion}{$v_{\text{current}}^s\left(t\right), q_{\text{current}}\left(t\right), \tau - \mbox{traveltime}$, $\theta$}
			\Comment{Obtain the successor node.}
			\State traveltime $\gets$ traveltime $+$ \textbf{random sample of} $T_{\text{successor, current}}$
			\Comment{Update the travel time.}
			\State \textbf{add} successor \textbf{to} $N_{\mathrm{V}}$
			\Comment{Extend the known subgraph.}
			\State current $\gets$ successor \Comment{Make the step to the successor.}
			\State {\bf add} $\left(\text{current, traveltime}\right)$ {\bf to} steps
		\EndWhile
		\State {\bf return} steps
	\EndFunction
\end{algorithmic}
\caption{Our decentralized routing algorithm (which builds on the iterative approximation scheme developed by Fan and Nie \cite{Fan2006Optimal}). The input parameters are a network $G$, an origin node, a target node, a time budget $\tau$, a CDF threshold $\theta$, and a \textsc{Criterion} to identify successor nodes.
}
\label{alg:algorithm}
\end{algorithm*}


\section{Mixture of Convolutions Versus Convolution of Mixtures} \label{app:conv}

Let $F=\{f_i\left(x\right)\}$ and $G=\{g_j\left(y\right)\}$ be two finite sets of probability density functions (PDFs), and let the mixtures of the elements of the sets be given by
\begin{align*}
	\bar{f}\left(x\right) &= \sum_i \omega_{f_i} f_i\left(x\right)\,, \\
	\bar{g}\left(y\right) &= \sum_j \omega_{g_j} g_j\left(y\right)\,,
\end{align*}
where $\omega_{f_i}$ and $\omega_{g_j}$ are, respectively, the independent weights associated with the elements of $F$ and $G$. Taking the mixture after performing the convolution of the elements of $F$ and $G$ gives
\begin{align*}
	\sum_{ij}\omega_{f_i} \omega_{g_j} \left(f_i \ast g_j\right)\left(z\right)
	&=\int_0^z dx\, \sum_{ij}\omega_{f_i} f_i\left(z-x\right) \omega_{g_j} g_j\left(x\right) \\
	&=\int_0^z dx\, \bar{f}\left(z-x\right) \bar{g}\left(x\right) \\
	&=\left(\bar{f}\ast\bar{g}\right)\left(z\right)\,.
\end{align*}
Therefore, as long as the assumption of independent weights holds, it follows that mixing the result of a convolution is equivalent to taking the convolution of two mixtures.

Consider $b$ sets of probability distributions $\left\{F_1,\ldots,F_b\right\}$. Let each set $F_i$ have $c_i$ elements. Carrying out the convolutions of all pairs of probability distributions in the sets first and taking the mixture afterwards requires $\left(\prod_{i=1}^b c_i\right)$ convolutions and additions. However, carrying out the mixtures first and then performing the convolutions requires $b$ convolutions and $\left(\sum_{i=1}^b c_i\right)$ additions. It is thus much more efficient computationally to compute the mixtures first and subsequently perform the convolutions.



\end{document}